\documentclass{aa}  
\usepackage{txfonts}
\usepackage{stfloats}
\usepackage{xcolor}
\usepackage{float}
\usepackage{afterpage}

\def\starlight{{\sc Starlight}}
\def\fado{{\sc FADO}}
\newcommand{\hchimi}{{\textsc{HII-CHI-mistry}}}
\def\sex{{\sc SExtractor}}
\def\steckmap{{\sc STECKMAP}}

\newcommand{\sbb}{mag/$\sq\arcsec$}
\def\mstar{${\cal M}_{\star}$}
\def\mstotal{${\cal M}_{\star,\textrm{T}}$}
\def\tsstar{$\Sigma_{\star}$}
\def\mssdss{${\cal M}_{\star,\textrm{1.5}}$}

\begin{document}

% =============================================================================================
\title{Characterisation of the stellar content of SDSS EELGs through self-consistent spectral modelling}
% =============================================================================================
   \author{
          Iris Breda \inst{\ref{IAA-CSIC}}
          \and          
          Jos\'e M. Vilchez \inst{\ref{IAA-CSIC}}
          \and          
          Polychronis Papaderos \inst{\ref{IA-FCiencias},\ref{ULisbon},\ref{UPorto}}
          \and          
          Leandro Cardoso \inst{\ref{UPorto}}
          \and          
          Ricardo O. Amorin \inst{\ref{ChileI},\ref{ChileII}}
          \and          
          Antonio Arroyo-Polonio \inst{\ref{IAA-CSIC}}
          \and                   
          Jorge Iglesias-P\'aramo \inst{\ref{IAA-CSIC}}
          \and          
          Carolina Kehrig \inst{\ref{IAA-CSIC}}
          \and          
          Enrique P\'erez-Montero \inst{\ref{IAA-CSIC}}
          }

\institute{Instituto de Astrof\'{i}sica de Andaluc\'{i}a, 
Glorieta de la Astronom\'{i}a, s/n, 18008 Granada, Spain \label{IAA-CSIC}
\and
Instituto de Astrofísica e Ciências do Espaço, Universidade de Lisboa, OAL, Tapada da Ajuda, PT1349-018 Lisboa, Portugal 
\label{IA-FCiencias}
\and 
Departamento de Física, Faculdade de Ciências da Universidade de Lisboa, Edifício C8, Campo Grande, PT1749-016 Lisboa, Portugal \label{ULisbon}  
\and 
1 Instituto de Astrofísica e Ciências do Espaço, Universidade do Porto, CAUP, Rua das Estrelas, PT4150-762 Porto, Portugal \label{UPorto} 
\and
Departamento de Astronom\'ia, Universidad de La Serena, Av. Juan Cisternas 1200 Norte, La Serena, Chile \label{ChileI}
\and
Instituto de Investigaci\'on Multidisciplinar en Ciencia y Tecnolog\'ia, Universidad de La Serena, Ra\'ul Bitr\'an 1305, La Serena, Chile \label{ChileII}
\\
             \email{breda@iaa.es}
}

   \date{Received ???; accepted ???}

\abstract{Extreme emission line galaxies (EELGs) are a notable galaxy genus, ultimately being regarded as local prototypes of early galaxies at the cosmic noon. Robust characterisation of their stellar content, however, is hindered by the exceptionally high nebular emission present in their optical spectroscopic data. This study is dedicated into recovering the stellar properties of a sample of 414 EELGs as observed by the SDSS Survey. Such is achieved by means of the spectral synthesis code \fado, which self-consistently considers the stellar and nebular emission in an optical spectrum. Additionally, a comparative analysis was carried on, by further processing the EELGs sample with the purely stellar spectral synthesis code \starlight, and by extending the analysis to a sample of 697 normal star-forming galaxies, expected to be less affected by nebular contribution. We find that, for both galaxy samples, stellar mass and mean age estimates by \starlight\ are systematically biased towards higher values, and that an adequate determination of the physical and evolutionary properties of EELGs via spectral synthesis is only possible when nebular continuum emission is taken into account. Moreover, the differences between the two population synthesis codes can be ascribed to the degree of star-formation activity through the specific star-formation rate and the sum of the flux of the most prominent emission lines. As expected, on the basis of the theoretical framework, our results emphasise the importance of considering the nebular emission while performing spectral synthesis, even for galaxies hosting typical levels of star-formation activity.}

\maketitle
%________________________________________________________________

\parskip = \baselineskip

\section{Introduction \label{intro}}

Throughout the many decades of research across the extra-galactic domain, several thousand galaxy specimens have been discovered that display substantially elevated equivalent widths (EWs; between 100 and up to $\sim$ 2000 \AA) of particular emission lines concomitant of violent star-formation (SF) events, such as H$\alpha$, [OIII]$_{5007}$, [OII] and Ly-$\alpha$, the latter being found mainly at the high-z regime. These galaxies, designated as extreme emission line galaxies (EELGs), may also be classified as green pea galaxies \citep[GPGs; e.g.,][]{Car09, Amor10} or as blueberries \citep{Yan16}, depending on their redshift and colours. EELGs can also be included in additional classification schemes such as blue compact galaxies or blue compact dwarfs \citep[BCGs/BCDs; e.g.,][]{KunSar86,Cai01,Pap96,ThuMar81,LooThu86,Rev07}, or ELdots \citep{Bek15}. The latter galaxy groups, however, are defined solely according to photometric properties, and include galaxies that are not EELGs. In addition, Wolf-Rayet features \citep[e.g.,][]{Scha99, Amor12} and nebular HeII emitters \citep[e.g.,][]{Keh18,Fer21} are found among EELGs. These are considered well-preserved relics alluding to the high-z Universe, where, in contrast with what is observed in the nearby Universe, such extreme events were increasingly more prevalent. Chiefly in virtue of their faint surface brightness levels, the true nature of these galaxies remains a mystery, with numerous fundamental questions yet to be addressed. 

Aside from conventionally being regarded as windows to cosmic history, emulating the conditions observed in high-z galaxies at the peak of SF, EELGs harbour a substantial number of young, massive star forming complexes which produce a considerable amount of photoionising radiation, successively being absorbed by the metal-poor gas existing in their surroundings \citep[e.g.,][]{Rav20}. Considering the abnormally intensive SF events and the ensuing high number of super-novae explosions, the gas content is expected to present an irregular, punctured distribution from where hard ionising radiation may escape \citep[e.g.,][]{Fle19A,Ber06,Izo16,Per20}, making these galaxies perfect laboratories for the exploration of Ly-continuum escape. An additionally important remark with respect to these atypical galaxies is that these are consensually accepted as key contributors to the reionisation of the Universe \citep[e.g.,][]{Wise14,MadHaa15,Sta16,Nai21,Mat21}, and as the prime building blocks of more massive, local galaxies \citep[e.g.,][]{PreSch74}. EELGs are amongst the most chemically poor extra-galactic objects in the local Universe, in respect to both their stellar and gas content \citep[e.g.,][]{Rosa07,JasOey13,APV10,Iso21,Koj21}. Hereupon, a comprehensive study of this galaxy genus, specially at lower redshifts, offers a unique opportunity to examine in detail the specifics of pristine environments, and how in these conditions processes such as SF, chemical enrichment or mass growth, unfold.

Acknowledging how critical is the role that these galaxies play in the broader context of galaxy formation and evolution emphasises the importance of cultivating a holistic understanding of the subject. In this regard, there is an abundant number of studies that collectively aid to outline a coherent picture of the inherent properties of such galaxies across time, shedding some light upon the topic. A key detail for establishing a comprehensive and consistent framework relies on the assembly of a substantial number of exemplars across redshift. Predominantly through the exploration of deep narrow- or broad-band photometry, several studies have contributed in this regard \citep[e.g.,][]{Str09,Wel11,Amor15,Atek11,Smit14}. However, the information that can be extracted from this data type is rather limited, with the ideal being to acquire spectroscopic information for an equivalently high number of galaxies. \citet{EPM21} (hereafter EPM21) describes the assembly of such a sample. By scanning approximately one million galaxies in the SDSS-DR7 \citep{Sanc10} with apparent magnitude brighter than 17.8 in the $r$ band through the Automated Spectroscopic K-means-based (ASK) classification scheme, the authors identified approximately 2000 EELGs at 0 $\lesssim$ $z$ $\lesssim$ 0.49 with associated SDSS spectroscopic data. The work here presented exploits the aforementioned galaxy sample, emerging as a complementary study.

The assembly of such samples has been instrumental for the assessment of the physical properties of these galaxies. For instance, several studies have significantly contributed on the characterisation of the gas content of EELGs (in addition to the review by \citealt{KunOst00}, see \citealt{Amor15}, EPM21, and \citealt{Amor10}) where the analysis of high-quality, spectroscopic data through a variety of methods \citep[e.g., direct t$_{e}$ method, strong line methods and the code \hchimi,][]{EPM14} revealed low  total oxygen abundances (7.3 -- 7.7 $\; \lesssim \;$ 12 + log(O/H) $\; \lesssim \;$ 8.5 -- 8.6) and equally low nitrogen-to-oxygen ratios (-1.8 $ \lesssim$ log(N/O) $\lesssim$ -0.8). The prior study additionally concluded that these display significantly compact morphologies, sharing many characteristics with high-redshift analogues, and that these objects are likely to be in a transient phase of their early evolution, currently building a significant amount of their stellar mass. 

As far as the gas kinematics in EELGs is concerned, previous research has demonstrated high turbulence and rapidly evolving scenarios. For instance, 
by investigating a sample of 22 EELGs at 1.3 $< z <$ 2.3 through near-infrared spectroscopy, \citet{Mas14} concluded that, at these redshifts, these low-mass, low-metallicity systems display velocity dispersions of the order of 50 km/s which, to attain stability against SF, implies a gas fraction above two-thirds. The authors add that the ongoing star-formation activity is expected to endure solely for $\sim$50 Myr, rapidly assembling stellar mass and decreasing the specific star-formation rate (sSRF) on a time-scale that is shorter than that estimated for the gas depletion ($\sim$100 Myr). Further evidence is provided by \citet{Bos19} where, upon inspection of GMOS-IFU data of one EELG, three individual kinematic components were identified; a rotating component that the authors correlate with a disk-like feature, a violent, extensive outflow and an intermediate component that is considered to be a turbulent mixing layer previously identified in local starburst galaxies. Further evidence is provided by \citet{Hog20}, who analysed high-dispersion long-slit data. These results once more highlight the underlying complex nature of these galaxy systems.

In respect to their stellar content, it is documented that a vast number of BCDs possess an old underlying stellar component, in addition to the young stellar complexes that recently emerged as a result of the ongoing, intensive SF activity \citep[e.g.,][]{Pap96,Sch99,Cai02,Amor12}. Despite the fact that there is not a substantial body of knowledge describing the stellar populations of EELGs through redshift, complementary studies have been conducted, being mostly based on spectro-photometric evolutionary synthesis such as spectral energy distribution (SED) fitting \citep{Jan17,Cohn18}. It is critical however, to standout the many considerable biases and degeneracies inherent to SED fitting techniques \citep[see e.g.,][]{Wal11}. In this regard, and with all their shortcomings \citep[see][for a review]{GomPap17}, population spectral synthesis (PSS) techniques are by far the most reliable method for the challenging task of accurately characterising stellar populations in galaxies. Nevertheless, as most of the available PSS codes are purely stellar, that is, they do not consider the nebular continuum contribution, these are not suited for the study of star-forming galaxies. For EELGs in particular, which register especially high nebular contributions, the use of conventional PSS tools would introduce significant biases, impeding the accurate estimation of star-formation and chemical enrichment histories (SFH and CEH, respectively). In this respect, \fado\ \citep{GomPap17,Car19,Pap21}, with its unique capability of self-consistently considering the nebular component whilst fitting the stellar templates to the observed optical spectrum, is the only PSS tool that is qualified to produce sensitive results (it was recently employed to characterise the stellar content of three GPGs in the work by \citealt{Fer21}).

The primary goal of the present study is to accurately characterise the stellar content of a representative sample of EELGs, by processing with \fado\ a subsample of approximately 400 galaxies from EPM21. The secondary goal is to evaluate how the estimated stellar properties, such as stellar mass and mean stellar age, vary when considering full self-consistent or purely stellar spectral synthesis tools. To reach this goal, extra runs with the purely stellar PSS \starlight\ \citep{Cid05} were performed. 
Finally, we aim to compare EELGs and normal SF galaxies. This supplementary analysis allows for the identification of the physical properties compelling the observed differences, and to quantify the extent of these differences in the two galaxy families.
The article is organised as follows: in Sect. \ref{meth} we introduce the samples to be analysed and the adopted methodology, and in Sect. \ref{res} we discuss the obtained results, simultaneously exploring the causes for the disagreement between the two PSS codes and the differences between EELGs and normal SF galaxies. Finally, Sect. \ref{conc} summarises the main results obtained in the course of this analysis.
 
% I acknowledge that FADO, with its unique capability of considering the nebular component in a self-consistent manner, would be of crucial importance (even considering possible Ly continuum escape and gas-stars non co-spatiality, I believe that its results would be remarkably better than the same obtained with Starlight - or its modified version for the processing of photospectra - or any available SED fitting tool). However, the data provided by the miniJPAS Survey and the data that will follow, has a spectral resolution of about R~60, which I believe is not enough to be processed by the publicly available version of FADO. Hereupon, I'm taking the liberty of proposing a possible collaboration for the development of a FADO version that is capable of dealing with such low resolution spectra. Additionally, in case that you consider that such joint project is impracticable, regardless of the reason, I'm open to any suggestions that you consider meaningful for the accurate characterization of the stellar content of extreme emission line galaxies. I'll also be available for any future discussion or questions that might arise.

\section{Sample description and data analysis \label{meth}}

The present work is subdivided into three stages, the first pertaining to isolating the best candidates from the original sample of EPM21, along with assembling a control sample of normal SF galaxies, the second referring to the spectral modelling of the SDSS spectra, and finally correction of the obtained physical properties for aperture effects.
 
% ::::::::::::::::::::::::::::::::::::::::::::::::::::::::::::

\begin{figure*}[t]
\centering
\includegraphics[width=1\linewidth]{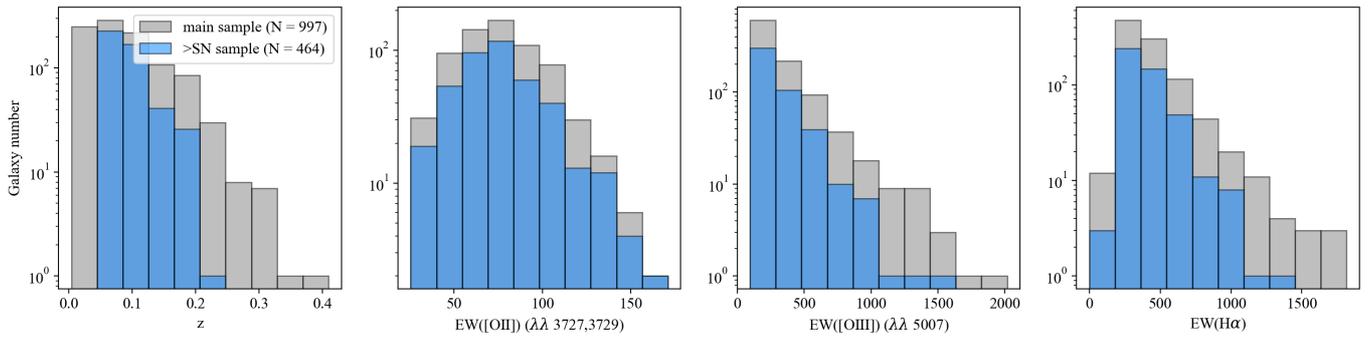}
\caption{Redshift and EW ([OII] ($\lambda\lambda$ 3727,3729), [OIII] ($\lambda\lambda$ 5007) and H$\alpha$) distributions of both the main EELG sample (grey) and the sub-sample with (S/N)$_{6390:6490}$ > 8 and to which the observed spectrum encompasses the Balmer jump (blue).}
\label{hist}
\end{figure*}

\subsection{EELG sample \label{samp}}
As briefly aforementioned, the galaxy sample here analysed was identified in EPM21 by employing the ASK classification scheme on the SDSS-DR7 data-base, composed by roughly 10$^{6}$ galaxies with apparent magnitude $<$ 17.8 in the $r$ band (the reader is referred to EPM21 for additional details on sample acquisition and specifics on ASK classification). The sample was identified by selecting galaxies that display an EW(H$\beta$) $>$ 30\AA\ and an EW([OIII]) $>$ 100\AA, yielding 1969 objects with a redshift distribution between 0 $<$ z $<$ 0.49. However, the execution of this study requires the fulfilment of certain additional conditions: 
\begin{itemize}
   \item[i.] To have available SDSS photometric data. The photometric analysis provides the means of visually inspecting the data thus confirming a) the galaxy morphology and b) the galaxy fraction covered by the fibre. In addition, it permits to estimate the radial extent of the galaxy, essential to correct from aperture effects. From the original 1969 objects of the starting sample, 1740 contain photometric information.
	\item[ii.] After visual inspection, cometary and tadpole galaxies to which the fibre did not incorporate a significant portion of the galaxy's radial extent and highly star-forming nodes in spiral galaxies, galaxy mergers or low surface brightness (LSB) galaxies were excluded from the main sample, thereafter listing 1119 objects.
	\item[iii.] By applying \sex\ \citep{sex} to the photometric $r$ frames, aperture photometry was performed, permitting to recover total radial extent R$_{\rm T}$ (i.e., the galactocentric radius at which the surface brightness level equalizes the background), effective radius R$_{\rm eff}$ and the galaxy's flux fraction integrated within the SDSS aperture (f$_{\rm ap}$). Comparison  with the photometric information provided by the SDSS-DR12 Photometric catalogue \citep{Ala15} has revealed an overall good agreement except for 10 galaxies ($\sim$1\%) where the latter provides significantly over-estimated sizes. To assure that the fibre does not encompass only a residual amount of the total galaxy light, galaxies with f$_{\rm ap}$ < 30 \% were excluded, thus obtaining the main sample of 997 sources.
	\item[iv.] Considering the study by \citet{Pap21}, where it was established a consistent dependence of the quality of the fits on the signal-to-noise ratio (S/N), and the dependence of the inclusion/exclusion of the Balmer jump at 3646\AA\ (and Paschen at 8207\AA\ for galaxies at  $z$ $\lesssim$ 0.12) on the resulting best-fitting model, it was selected a sub-sample of galaxies with (S/N) > 8 (measured at the restframe emission-line-free pseudo-continuum between 6390\AA\ and 6490\AA) and a rest-frame wavelength starting at < 3640\AA, listing 464 galaxies.
\end{itemize}

\begin{figure*}[h!]
\centering
\includegraphics[width=0.8\linewidth]{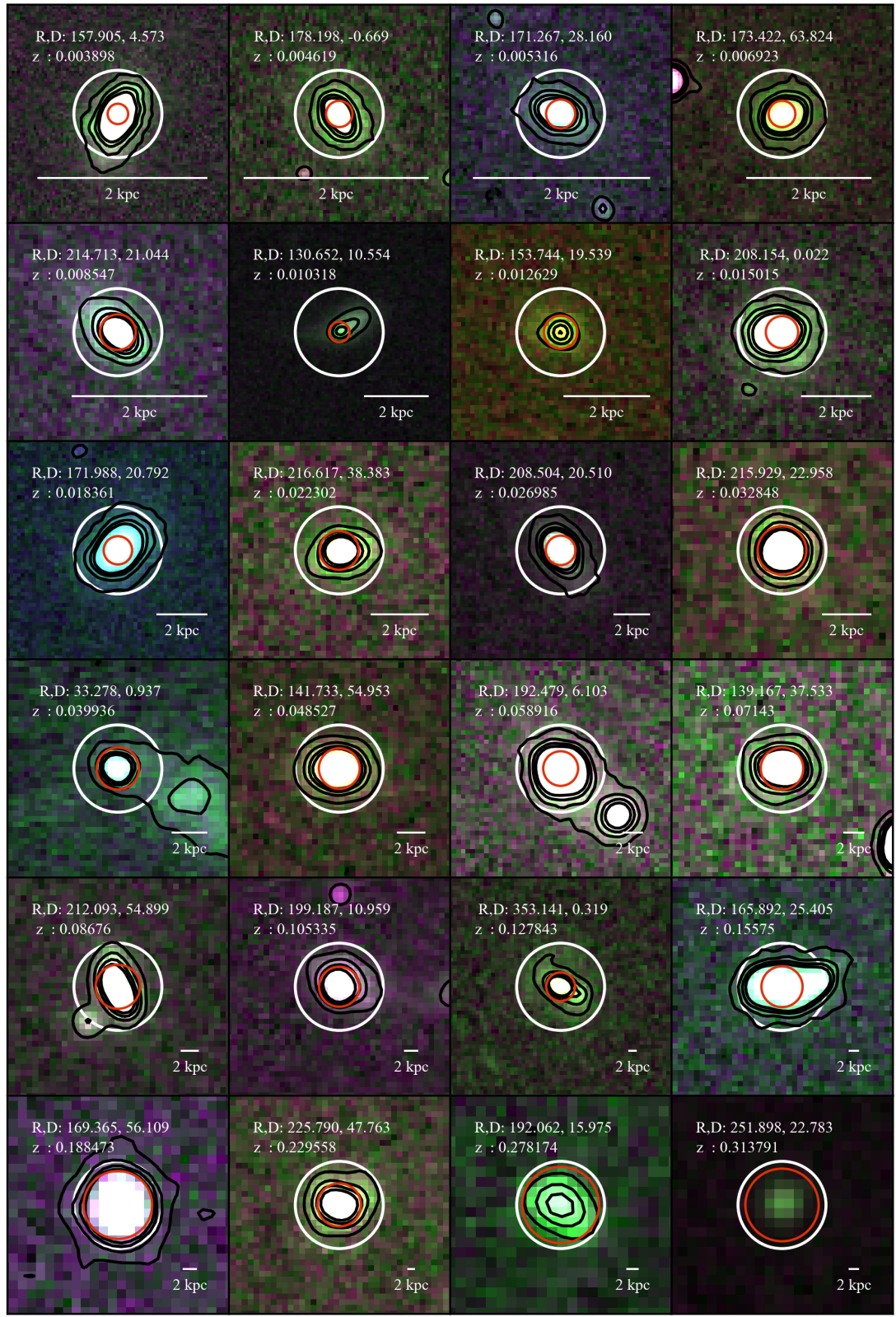}
\caption{True-colour SDSS images of 24 randomly selected EELGs at different redshifts. It is shown the RA and DEC, redshift and a scale-bar depicting 2 kpc. The black contours show the galaxy morphology by highlighting the $r$ band isophotes, the white circle portrays the estimated galactic radial extent and the red circle represents the SDSS aperture ($\protect\diameter$ = 3\arcsec). The galaxies are ordered from top to bottom and from left to right, in ascending redshift order.}
\label{tc}
\end{figure*}

Figure \ref{hist} illustrates the distribution of the redshift and of some typically high EWs for the main sample, encompassing all the EELGs from EPM21 that satisfy the previously mentioned criteria (i.e., have available SDSS photometric data, are individual, compact galaxies and have f$_{\rm ap}$ > 30\%), and the subsample consisting of the EELGs with (S/N)$_{6390:6490}$ > 8, and to which the Balmer jump is present in the observed spectrum. The first panel shows that the vast majority of the sampled galaxies have $z$ $\lesssim$ 0.2, reaching the peak of the distribution at 0.05 < $z$ < 0.1. With the exception of galaxies with $z$ < 0.05 and $z$ > 0.2, which are not well represented by the reduced, higher-S/N sample, both distributions are comparable. However, the redshift range is not sufficiently wide to conduct a sound exploration across time. The additional panels of Fig. \ref{hist} display the distribution of the EWs of several emission lines, namely, from left to right, the flux-weighted average of the [OII] doublet at 3727\r{A} and 3729\AA, the [OIII] at 5007\r{A} and H$\alpha$ at 6563\AA. It is apparent that the entire range of EWs is properly represented by the reduced, higher S/N sample. For illustrative purposes, 24 EELGs were uniformly selected across redshift, portraying some of the typical morphologies of the sampled EELGs across time, as  exemplified by Fig. \ref{tc}. Inspection of this figure reveals that although many EELGs display compact, circular morphologies, a significant fraction thereof are elongated, as apparent from the overlay contours.

\subsection{Control sample of normal SF galaxies \label{Csamp}}

For completeness, and aiming for a comparison study, the analysis was extended to a sample of normal SF galaxies. After identifying the plate, MJD and fibre of the ASK galaxies that conform to the defined selection criteria, spectroscopic SDSS data were retrieved from the SDSS DAS User Fiber List Submission Form\footnote{http://das.sdss.org/www/html/post\_fibers.html}. The selection was based on their BPT [NII]/H$\alpha$ versus [OIII]/H$\beta$ values \citep{BPT}, specifically by selecting the galaxies falling below the \citealt{Kauf03} demarcation line (corresponding to pure SF systems), with a (S/N) > 3 in the most relevant emission lines (i.e., [OIII], H$\alpha$, H$\beta$) and within the redshift range of 0.04 $\leq$ $z$ $\leq$ 0.1.

%\vspace{-0.7cm}

%\begin{figure*}[b]
%\centering
%\includegraphics[width=1\linewidth]{fig/fit_example.eps}
%\caption{Exemplary spectral modelling result for one of the EELGs of the sample . The left-hand-side panel shows the resulting fit as obtained by \fado\ and the right-hand-side the equivalent for \starlight. The stellar fit is shown in red and the nebular continuum fit (exclusively for \fado) in blue. The residuals are shown in black. In addition, the panel corresponding to the \starlight\ fit displays in light blue the applied mask for the emission lines.}
%\label{example_fit}
%\end{figure*}

Additionally, to secure the extraction of normal SF galaxies (i.e., excluding galaxies experiencing residual or violent SF events), the range intervals 10 $\leq$ EW([OIII]$_{5007}$) $\leq$ 100\AA, 20 $\leq$ EW(H$\alpha$) $\leq$ 100\AA\ and 3 $\leq$ EW(H$\beta$) $\leq$ 30\AA\ were selected. Subsequently, the obtained sample was cross-matched with the ASK catalogue \citep{AscSan11} and all galaxies with primary ASK class corresponding to EELGs were withdrawn. This approach resulted in the selection of 11832 objects, from which a sample of 1350 objects was extracted, by randomly selecting 150 galaxies from each of the 9 redshift bins. The EW distribution of the relevant emission lines in the final randomly selected sample was contrasted with the same distribution of the original sample of $\sim$12k objects, thus confirming that the reduced sample accurately represents the extended sample of normal SF galaxies. Comparatively with the EELGs sample, objects with (S/N) > 8 within the wavelength range between 6390\AA\ and 6490\AA\ were selected, resulting in a final control sample of 777 galaxies.

\begin{figure*}[b]
\centering
\includegraphics[width=1\linewidth]{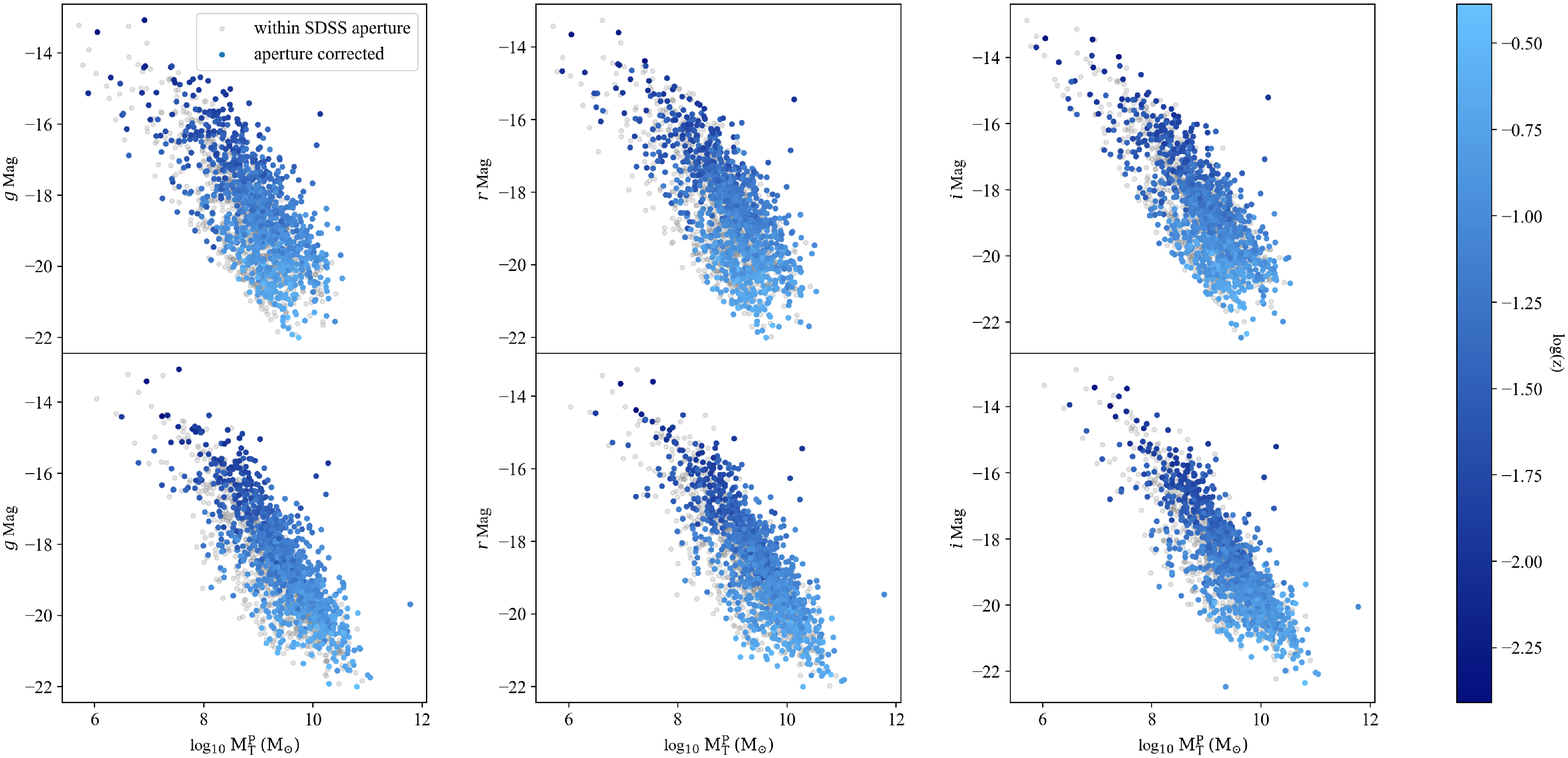}
\caption{Absolute magnitude in the $g$, $r$, and $i$ SDSS bands, from left to right, versus estimated present-day stellar mass for the EELG galaxy sample. The top panels display the stellar mass estimates as obtained by \fado\ and the bottom as obtained by \starlight. Grey circles illustrate the values obtained within the SDSS aperture and coloured circles display the corrected values. The colour-bar indicates the logarithm of the galaxy redshift.}
\label{mass}
\end{figure*}

% ::::::::::::::::::::::::::::::::::::::::::::::::::::::::::::
\subsection{Spectral modelling \label{mod}}
After de-redshifting the SDSS spectra (resolution $\lambda/\Delta\lambda \sim$2.5 at $\sim$3800\AA\ and $\lambda/\Delta\lambda\sim$3.6 at $\sim$9000\AA), rebinning to a step of 1\AA\ (for \starlight\ runs, it is recommended to provide the input spectrum de-redshifted and rebinned to 1\AA), and correcting for Galactic extinction \citep[following the extinction curve by][with R$_{V}$ = 3.1]{CCM89}, spectral modelling of both samples was performed by means of the PSS codes \fado\ and \starlight\ (as an integrity check, supplementary runs for the EELG sample were carried out with \fado\ purely stellar mode, \fado$\rm _{ST}$, that is, without considering the nebular contribution). As Balmer and Paschen jumps are prominent nebular features in SF galaxies (the latter being observed in SDSS galaxies at $z$ $\lesssim$ 0.12), and considering that the purely stellar code \starlight\ cannot account for such \citep[inclusion of these features would result in inaccurate fits, translating into stellar mass overestimation for younger ages, see][]{Car19}, \starlight\ runs were performed by not considering these spectral ranges (i.e., from 3670\AA\ to 8100\AA). As for \fado, the whole wavelength range was considered while adopting full-consistency mode, with the auxiliary \fado$\rm _{ST}$ runs being carried out within the fitting range adopted for \starlight. The main stellar library was designed, comprising 152 simple stellar population (SSP) spectra from \citet{BruCha03}, for 38 ages between 1 Myr and 13 Gyr for four stellar metallicities (0.05, 0.2, 0.4 and 1.0 $Z_{\odot}$), referring to a Salpeter initial mass function and Padova 2000 tracks. The individual SSP libraries for each galaxy were derived from the main library, by removing a varying number of SSPs with corresponding ages that are older than the age of the Universe at the $z$ of the galaxy under study. 
%Visual representation of an exemplary fit can be seen in Fig. \ref{example_fit}, for the galaxy at RA 147.59718 \& DEC 0.70813296 which was randomly selected from the EELG sample. The left-hand side illustrates the spectral modelling result as obtained with \fado, where both the resulting best stellar fit (in red) and nebular continuum (in blue) can be identified. The right-hand side illustrates the obtained best stellar fit with \starlight\ and the adopted emission line mask, which covers about 50 emission lines in the optical spectral range. The residuals can be appreciated on the bottom panels. 

In addition to the stellar properties pertaining to spectral modelling, \fado\ provides fluxes and EWs of a comprehensive array of emission lines. The obtained fluxes were corrected for the effects of intrinsic extinction using the Balmer decrement as measured by the H$\alpha$/H$\beta$ ratio, following the prescription proposed by \citet{Vogt03}, equations A11 and A12, with R$_{V}^{A}$ = 4.5 and R$_{\alpha,\beta}$ = 2.86.
Subsequently, star-formation rates (SFRs) and specific star-formation rates (sSFRs) were derived by following the prescription by \citet{Ken09}:

\vspace{-0.8cm}
\begin{equation}
\mathrm{SFR \, (M_{\odot}/yr) = 5.5 \times 10^{-42} \; L(H\alpha) \, (ergs/s),}
\end{equation}

where L(H$\alpha$) represents the dust-corrected H$\alpha$ luminosity. The sSFR is the SFR divided by the estimated stellar mass. Considering that this prescription has the underlying assumption of a constant dust-heating stellar population, which might substantially underestimate the true SFR in young stellar systems \citep{Ken09,WeiFri01}, the additional prescription by \citet{salva} (Eq. 6) was used. Both are in excellent agreement, not introducing any substantial alteration in the forthcoming correlations. However, it is important to note that SFR calibrations predicated on the H$\alpha$ luminosity enclose the assumptions of a) a continuous SF at a constant SFR for a minimum period of 70 to 80 Myr, and b) stellar populations of solar metallicity, which might not hold true. For instance, in the case that the stellar metallicity is strongly subsolar, the Lyman continuum photon production rate per unit mass is by a factor 3 to 4 times higher than for solar-metallicity stellar populations. This would imply a strong overestimation (underestimation) of SFRs (specific SFRs) based on H$\alpha$ luminosities. In addition, throughout this work it is considered a null escape fraction of ionising photons. Accordingly, the estimated SFRs might be regarded as first-order estimates.

% ::::::::::::::::::::::::::::::::::::::::::::::::::::::::::::

\begin{figure}[h]
\includegraphics[width=\columnwidth]{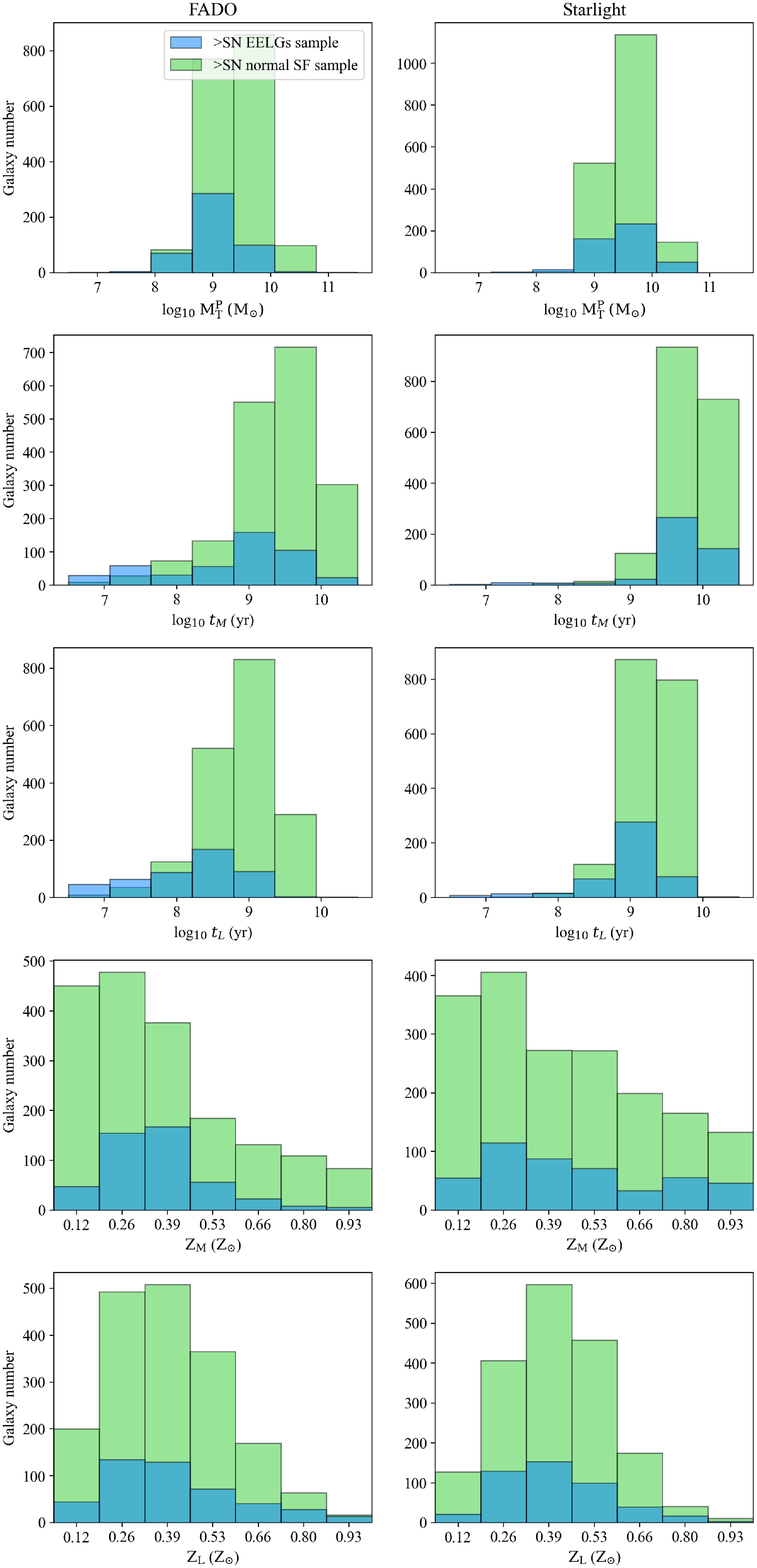}
\caption{Series of histograms summarising the main results obtained for both samples (EELGs in blue and normal SF galaxies in green), as obtained by \fado\ (left column) and \starlight\ (right column). From top to bottom, it displays, the distribution of the present-day stellar mass, the mass-weighted mean stellar age, the luminosity-weighted mean stellar age, the mass-weighted mean stellar metallicity, and the luminosity-weighted mean stellar metallicity.}
\label{sum}
\end{figure}

\subsection{Aperture corrections \label{ap_cor}}

The main goal of the present study is to characterise a sample of a considerable number of EELGs through systematic examination of single-fibre spectroscopic data. To this end, it is imperative to extrapolate the derived quantities (aperture based) to quantities that reflect the whole radial extent of the sampled galaxies. 

Aperture- and extinction-corrected H$\alpha$ luminosities and SFRs were obtained following the prescription by \citet{salva}, where it were derived aperture corrections for approximately 210,000 SDSS star-forming galaxies based on empirical H$\alpha$ growth curves extracted from the Calar Alto Legacy Integral Field Area (CALIFA) survey \citep{califa}. 

Concerning total stellar mass estimates, bearing in mind that spectroscopic information outside of the aperture is not available, a flawless correction for aperture effects is an unattainable concept. However, it is possible to obtain approximate estimates by assuming that these objects exhibit a constant mass-to-light (M/L) ratio, which is a reasonably valid assumption, considering that most of these objects are compact and uniform. In any case, it should be noted that, if the SF component (low-M/L ratio) is embedded within a more extended, older (higher-M/L ratio) stellar host, the total stellar mass would be underestimated, while the sSFR would be strongly overestimated. In addition, postulating a constant M/L ratio denotes that the previously estimated mean stellar age and metallicities within the fibre are similar throughout the whole galaxy extent.

%Concerning total stellar mass estimates, although flawless correction from aperture effects is an unattainable concept, regarding that spectroscopic information outward the aperture is not available, it is possible to obtain adequate results by assuming that these objects exhibit a constant mass-to-light (M/L) ratio. Furthermore, considering the compact nature of these objects, such assumption is reasonably valid. In addition, postulating a constant M/L ratio denotes that the previously estimated mean stellar age and metallicities within the fiber are equivalent though the whole galaxy extent.

The fair assumption that the surface brightness profile ($\mu$; \sbb) obtained in the spectral region where emission lines are absent or display a minimal contribution is a suitable proxy of the stellar surface density (\tsstar; i.e., the stellar mass \mstar, over the projected area, \mstar/$\rm \pi R^{2}$) provided the means to extrapolate outwards the aperture the stellar mass previously derived through spectral modelling of the SDSS spectra. Summarising the complete procedure:

\begin{itemize}
   \item[i.] For each galaxy, $\mu$ for the three SDSS bands $g$, $r$ and $i$ were extracted from the photometric frames (after sky modelling and subtraction) by fitting ellipses to the galactic isophotes.
   \item[ii.] As expected, the SDSS filters cover different intervals of the rest-frame spectrum according to the redshift of the source. Such can amplify or diminish colour gradients between the SF component and the surrounding LSB host, depending on redshift and the colour considered \citep{PapOst12}.
   %Figure \ref{FTC} depicts the filter coverage as a function of the observed $z$. It is evident that, 
   For instance, for galaxies at $z$ = 0, the $i$ band covers a spectral region which encompasses minimal nebular contribution, as this bandpass in the most sensitive to the stellar surface density profile of the galaxy. At $z$ = 0.12 however, the $g$ band provides a better tracer of the stellar mass, given that the strong [OIII]$_{5007}$ falls in the lower-transmission edge of that filter. With these considerations, the appointed $\mu$ $\equiv$ \tsstar\ was extracted from the $i$ band for galaxies with $z$ $\leq$ 0.067, from the $r$ band for 0.067 < $z$ $\leq$ 0.105 and from the $g$ band for $z$ < 0.105.

   \item[iii.] The $\mu$ were converted to linear units of counts/$\sq\arcsec$ ($\mu_{\rm int}$), spline interpolated to a constant finer step and integrated within the radius of the fibre, 1.5'' such as,

\vspace{-0.5cm}
\begin{equation}
\rm I_{1.5} = 2 \pi $\tsstar$ \int_{0}^{1.5} \mu_{\rm int} \rm \; R \; dR,
\end{equation}

where R is the galactocentric radius.

   \item[iv.] The previously obtained $\rm I_{1.5}$ over the stellar mass within the SDSS aperture \mssdss, provides the factor f necessary to convert the integrated $\mu_{\rm int}$ (i.e., the total number of counts in the specific filter) into \mstar. As a result, the total stellar mass \mstotal\ is obtained by integrating the entire $\mu_{\rm int}$ (i.e., until R$_{\rm T}$), and multiplying by f in the following manner:

\vspace{-0.5cm}
\begin{equation}
$\mstotal$ = 2 \pi \rm \; f \; $\tsstar$ \int_{0}^{\rm R_{\rm T}} \mu_{\rm int} \rm \; R \; dR
\end{equation}

Application of this procedure to the stellar masses obtained with both \fado\ and \starlight\ resulted in the estimated total stellar mass for the two PSS codes.
\end{itemize}

\begin{figure*}[t]
\centering
\includegraphics[width=1\linewidth]{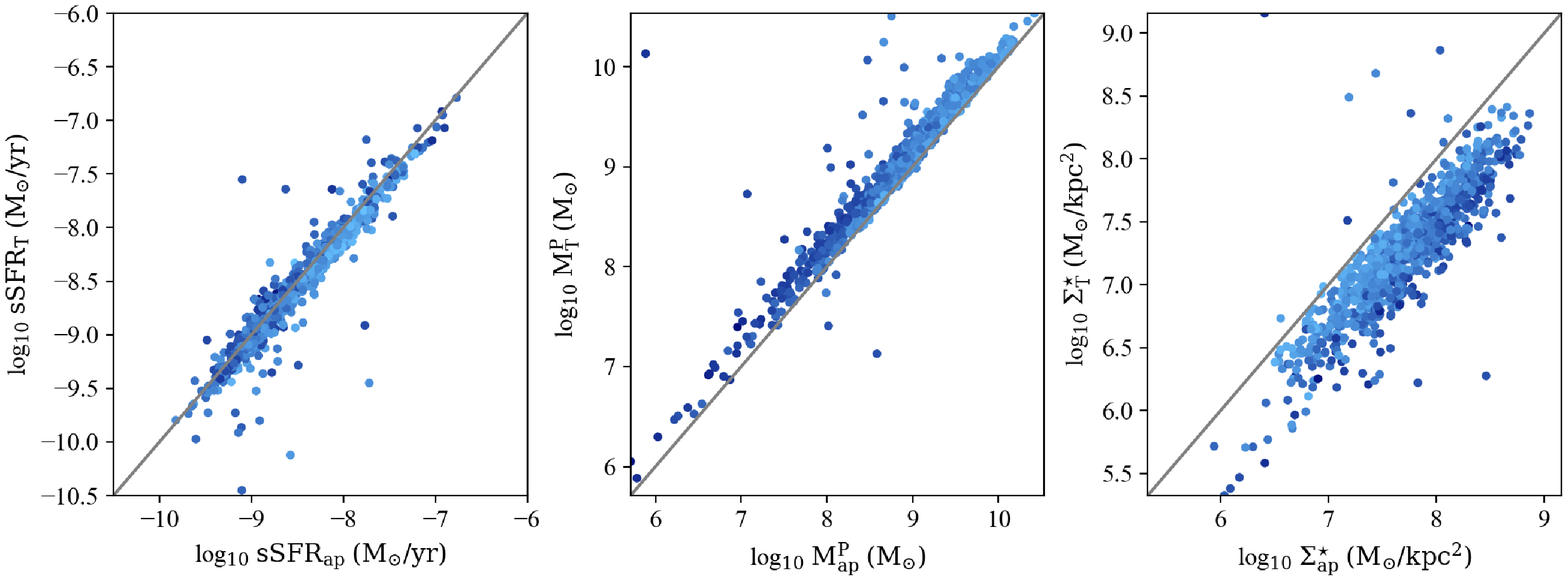}
\caption{Differences between aperture-corrected (x axis) and non-corrected quantities (y axis), such as sSFR (left-hand-side panel), present-day stellar mass as obtained by \fado\ (central panel) and stellar surface density (right-hand-side panel), for the EELG sample. Identity lines are represented in grey.
As in Fig. \ref{mass}, data points are coloured as specified by their redshift distribution.
}
\label{ap_difs}
\end{figure*}

Figure \ref{mass} exhibits the relation between absolute magnitude in the different SDSS optical bands and the derived stellar mass. It illustrates that the aperture-corrected stellar mass versus the total absolute magnitude within an aperture with radius R$_{\rm T}$ and the stellar mass versus the absolute magnitude within the SDSS aperture manifests an equivalent behaviour, supporting the soundness of the aforementioned approach. In addition, Fig. \ref{ap_difs} displays the relations between aperture-corrected and non-corrected quantities, namely for the sSFR (left-hand-side panel), present-day stellar mass as obtained by \fado\ (central panel) and stellar surface density (right-hand-side panel), for the EELG sample. While the sSFR does not exhibit significant variation, the mass and $\Sigma_{\star}$ display a clear shift towards higher masses and lower $\Sigma_{\star}$, respectively. The observed behaviour is expected, considering that for these galaxies the stellar mass fraction outside the fibre is typically low, whereas their radial dimension extends quite beyond the fibre, with an average galactic radius of $\sim$ 3.1'' $\pm$ 0.1, in the SDSS bands  $r$, $g$ and $i$.

Regarding the normal SF galaxies, in virtue of their considerable size and heterogeneous morphology and nature, no attempt was made to perform aperture corrections. Accordingly, the comparison study can only be carried out between non-corrected stellar properties, such as mass- and luminosity-weighted age and metallicity, and physical properties as estimated before aperture correction.

\section{Spectroscopic analysis}\label{res}

\subsection{Recovering fundamental correlations between galactic properties}\label{res1}

The EELGs and normal SF galaxy samples were processed as described in the previous section, providing a great amount of relevant results for both the stellar and gas components. However, considering that the gas-phase of these EELGs was extensively studied in EPM21, the present study focuses on their stellar properties (mainly mean stellar age and stellar mass) and on how both PSS codes here probed recover these properties, while contrasting with the same obtained for normal SF galaxies. 

\begin{figure*}[b]
\centering
\includegraphics[width=1\linewidth]{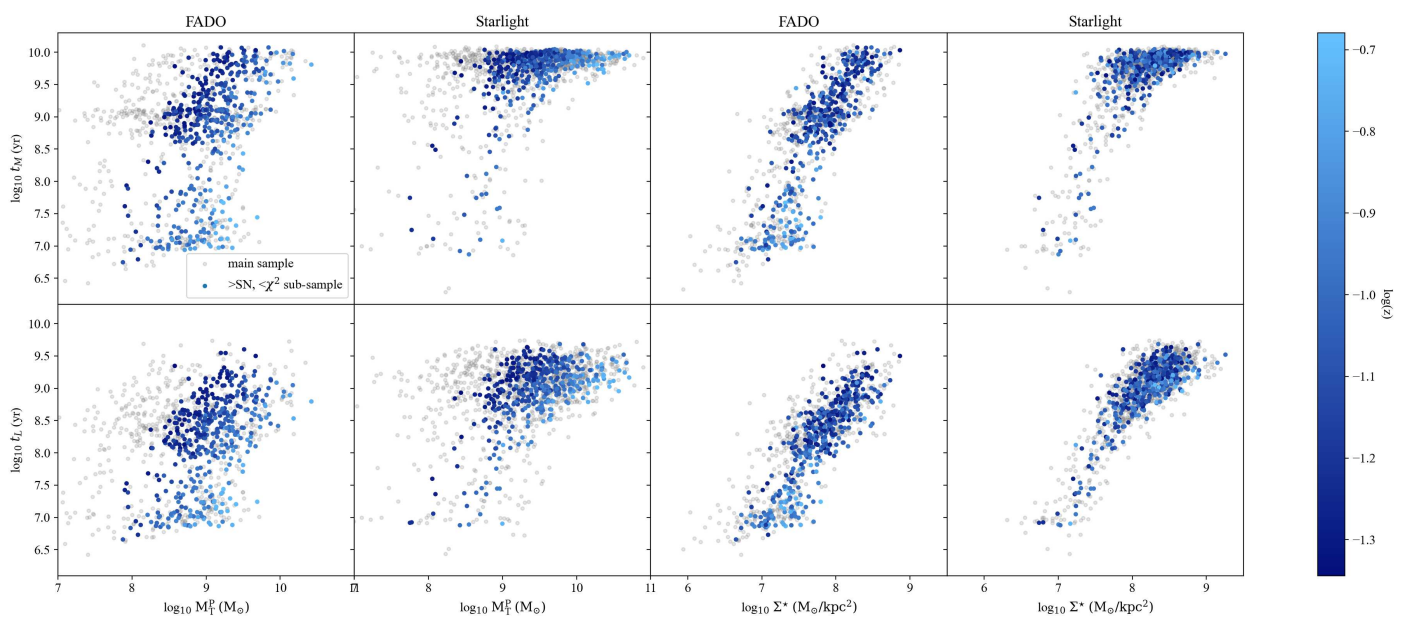}
\caption{Correlations between the mean stellar age (mass-weighted in the top row and luminosity-weighted in the bottom row) and the corrected total stellar mass (first four panels) and corrected surface stellar density (last four panels), as obtained by \fado\ (first and third columns) and \starlight\ (second and forth columns). Coloured circles depict the high-S/N EELG sample whereas grey circles depict the main EELG sample. The colours correspond to the redshift distribution of the sample, as given by the colour-bar displayed in the right-hand-side of the plot.}
\label{age_rel}
\end{figure*}

Both samples were once more curtailed by rejecting poor fits, that is, the galaxies for which the reduced $\chi^{2}$ of the fit by both PSS codes is higher than 3. The above resulted in 414 EELGs and 697 normal SF galaxies. Inspection of Fig. \ref{sum} reveals that, when comparing both samples, greater differences are found for the properties as obtained by \fado. This can be seen mostly for the logarithm of the present-day stellar mass, where the great majority of EELGs display values between 8.5 and 9.5 whereas the bin 9.5 to 10.5 is mainly populated by normal SF galaxies. Mass- and luminosity-weighted mean stellar age also reflects similar behaviour, considering that a substantial fraction of EELGs are significantly younger, as expected. \starlight, on the other hand, does not recover any fundamental difference between these two galaxy families. Bearing in mind that \starlight\ is a purely stellar code, it is incapable of disentangling between the stellar and nebular contributions. As a result, the nebular continuum generated by photoionisation of the gas by young stellar populations, which has a higher contribution in the blue wavelength range, is regarded as old, high-M/L ratio stellar populations. This effect, first pointed out by \citet{Izo11}, subsequently translates into over-estimating masses and mean stellar ages \citep[see][for a detailed analysis]{Car19}, as it is properly demonstrated by Fig. \ref{sum}, where both galaxy families are, on average, mainly regarded by \starlight\ as more massive, significantly older stellar systems.

\begin{figure*}[t]
\centering
\includegraphics[width=1\linewidth]{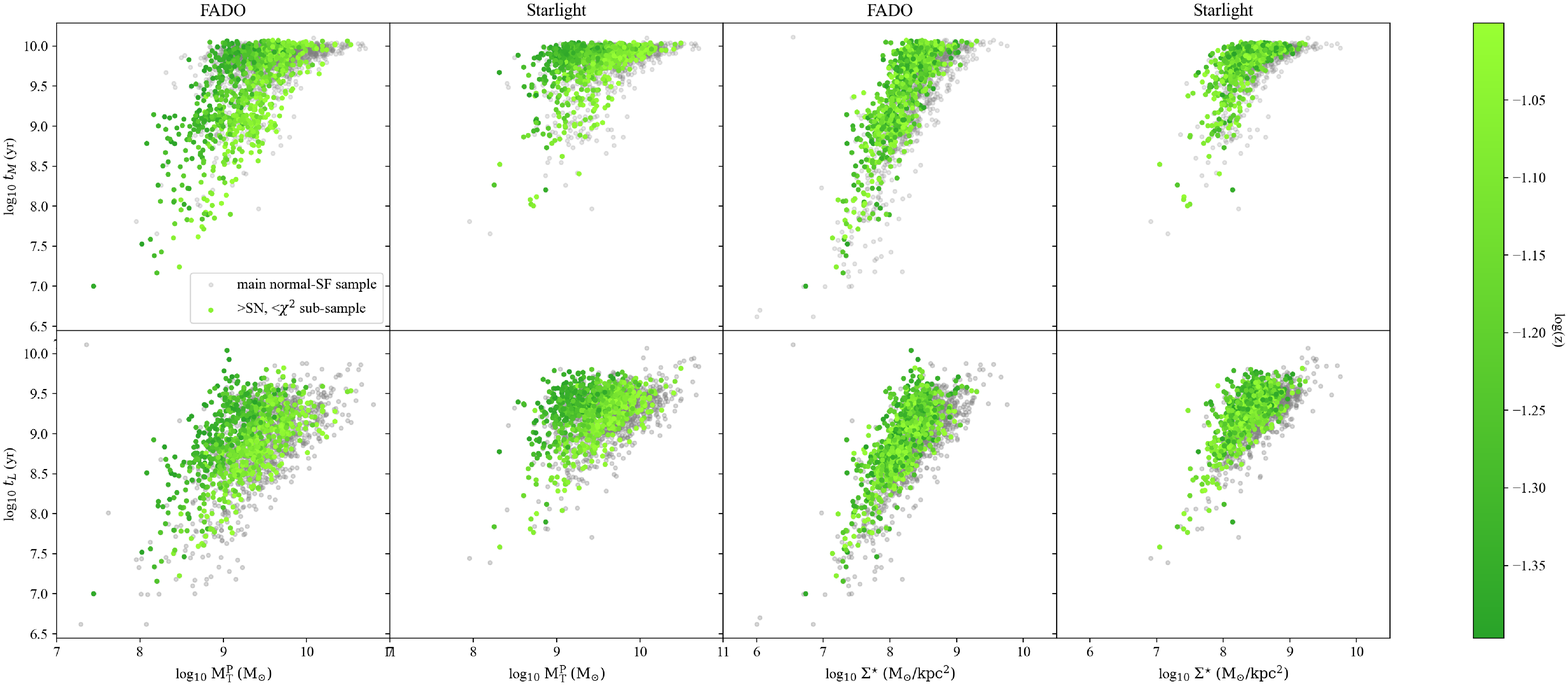}
\caption{Repetition of Fig. \ref{age_rel} for the normal SF sample, displaying non-corrected quantities (from aperture effects) for the present-day stellar mass and surface stellar density.}
\label{age_rel_normSF}
\end{figure*}

Figure \ref{age_rel} clearly demonstrates the previously reported phenomenon: whereas \fado\ recovers a positive trend between mass-weighted mean stellar age and total stellar mass (although highly scattered) and surface stellar density, \starlight's mass-weighted age determinations appear to be saturated, being primarily concentrated at the high-mass locus (although to a lesser degree, the same behaviour is seen for luminosity-weighted mean age determinations). The fact that \starlight\ is unable to recover these fundamental relations suggests that correct determination of the stellar properties in SF galaxies requires proper treatment of the nebular contribution, as supported by the theoretical framework.
%the importance of considering the nebular contribution in the spectral modeling of SF systems , simultaneously evidencing the high level of inaccuracy when not. 
Equally important is the fact that the same behaviour occurs for normal SF galaxies (although to a lesser extent, considering that these galaxies display lower levels of nebular contamination), as can be appreciated upon inspection of Fig. \ref{age_rel_normSF}. Once more, this piece of evidence suggests that purely stellar codes such as \starlight\ can be severely affected by typical SF activity. 
%not requiring extremely high values of nebular emission such as the ones observed in EELGs to .

\begin{figure}[h]
\centering
\includegraphics[width=1\linewidth]{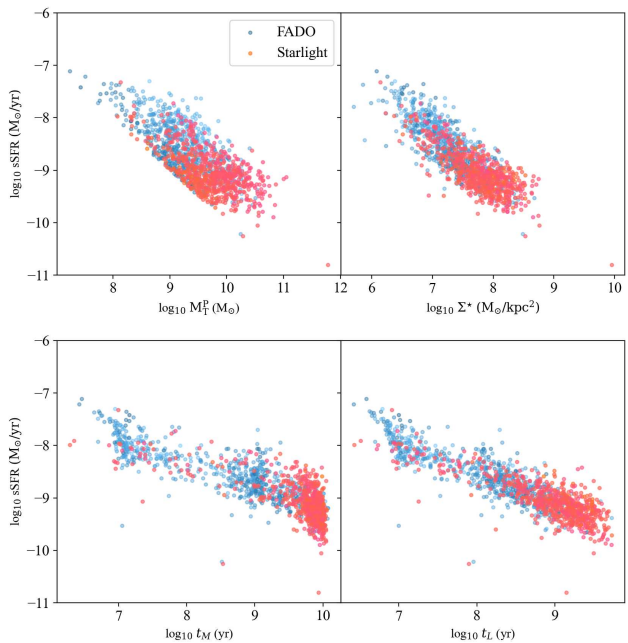}
\caption{On the top row it is displayed the logarithm of the aperture corrected sSRF versus the aperture-corrected stellar mass (left panel) and the aperture-corrected \tsstar (right panel). The bottom row displays the logarithm of the aperture-corrected sSFR versus the mass-weighted mean stellar age (left panel), followed by the same versus the luminosity-weighted mean stellar age (right panel).}
\label{sSFR_FD_SL}
\end{figure}

\begin{figure}[h]
\centering
\includegraphics[width=1\linewidth]{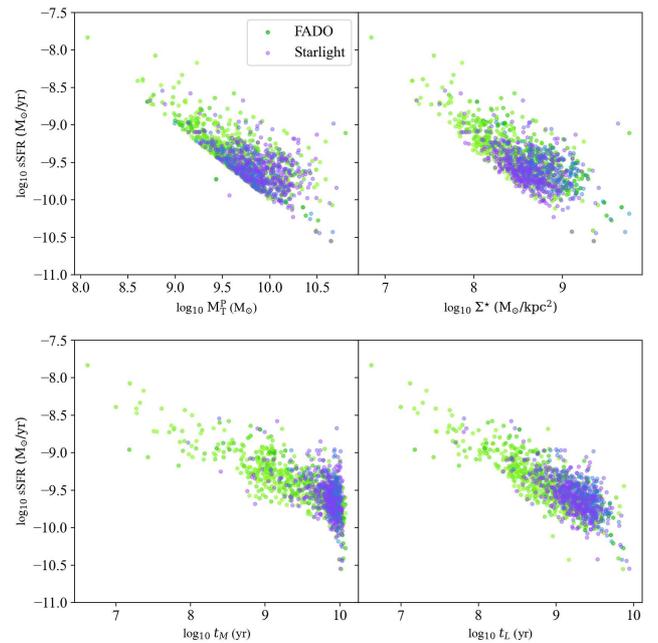}
\caption{Equivalent to Fig. \ref{sSFR_FD_SL} but for the normal SF galaxy sample. Contrary to the EELG sample, stellar mass and surface density estimates are not corrected from aperture effects.}
\label{sSFR_FD_SL_normSF}
\end{figure}

Additional evidence demonstrating the limited capability of a purely stellar code to soundly extract the physical properties of the stellar component of SF galaxies is provided by Figs. \ref{sSFR_FD_SL} and  \ref{sSFR_FD_SL_normSF} where it is displayed the relation between the sSFR and the total stellar mass, stellar surface density and mean stellar age for the main EELG sample. Although \starlight\ is capable of recovering the negative trends between sSFR and the total stellar mass and stellar surface density, the low-mass regime is significantly underpopulated, reflecting the tendency for the purely stellar code to provide enhanced mass estimates. This behaviour is also seen for the normal SF galaxies in Fig. \ref{sSFR_FD_SL_normSF}, which, alternatively, shows non-aperture-corrected quantities. Regarding the correlation between mean stellar age and sSFR, \starlight\ fits yield significantly high values for the majority of the sample, with both mass- and luminosity-weighted mean stellar age estimates displaying a severe saturation at the high end. Once more, the same behaviour is observed for the normal SF sample. In opposition, \fado\ recovers clear negative correlations between the sSFR and the aforementioned stellar properties, in agreement with previous studies \citep[e.g.,][]{Bri04,Noe07}.

\subsection{Inspecting the factors impacting the determination of the stellar properties}\label{res2}

The following stage is to assess the observed differences between spectral synthesis quantities (namely, stellar mass and mean stellar age) and identify the physical properties accountable for such contrast. The results outlined in the previous section strongly suggest that not accounting for the nebular contribution when fitting spectra of SF galaxies produces substantial biases towards higher ages and masses, ultimately concealing fundamental relations between galaxy properties. The results obtained by employing \fado$\rm _{ST}$ support the preceding conjecture.
%, however the difference in stellar mass and age is not as notable as when comparing with \starlight\ estimates (specifically, we obtained, for the logarithm of the present stellar mass for \fado\ $\rm log_{10}M^P_{\star}$ =  9.06, \fado$\rm _{ST}$ $\rm log_{10}M^P_{\star}$ =  9.30, \starlight\ $\rm log_{10}M^P_{\star}$ = 9.46; and for the logarithm of the mass-weighted stellar age for \fado\ $\rm log_{10}$ $t_{\rm M}$ = 8.75, \fado\ (ST mode) $\rm log_{10}$ $t_{\rm M}$ = 8.97, \starlight\ $\rm log_{10}$ $t_{\rm M}$ = 9.67). 

\begin{figure*}[t]
\centering
%\includegraphics[width=1\linewidth]{fig/comp_st_mode1.eps}
%\caption{Mass-weighted mean stellar age distribution with present-day stellar mass, as obtained by \starlight, \fado\ and \fado\ ST mode. The \textcolor{red}{left-hand side} panel compares the aforementioned distribution as obtained for \starlight\ and \fado\ ST mode, whereas the \textcolor{red}{right-hand side} panel compares the same as obtained for \fado\ and \fado\ ST mode. The histograms aid to discern the density of the different distributions.}
\includegraphics[width=1\linewidth]{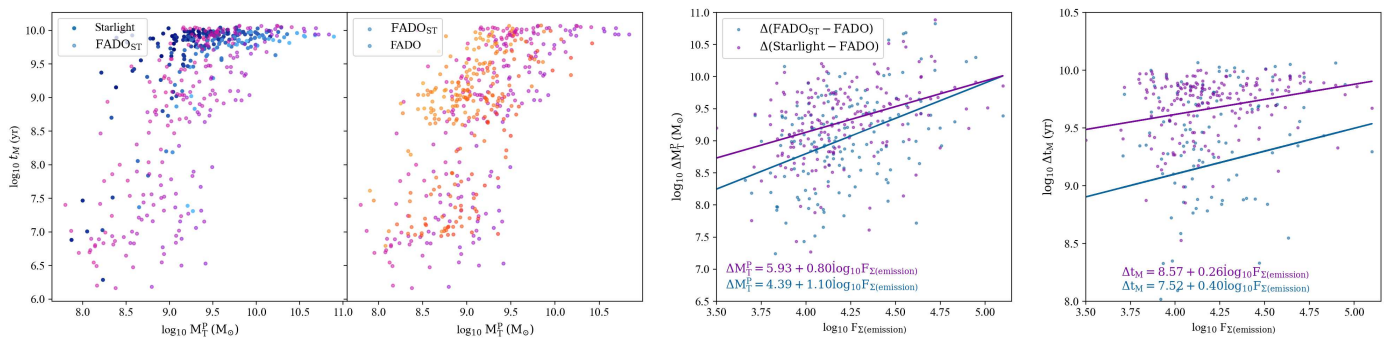}
\caption{Mass-weighted mean stellar age distribution with present-day stellar mass, as obtained by \starlight, \fado\ and \fado$_{\rm ST}$. The left-hand side panel compares the aforementioned distribution as obtained by \starlight\ and \fado$_{\rm ST}$, whereas the middle panel compares the same as obtained for \fado\ and \fado$_{\rm ST}$. The two right-hand side panels demonstrate how the normalised difference between the estimated stellar mass and mass-weighted stellar age (considering and ignoring the nebular contribution) relates to the logarithm of the flux associated with the sum of all emission lines, log$_{10}$F$_{\Sigma(\rm emission)}$, in units of 10$^{-16}$ erg/cm$^{2}$.s, which is a direct proxy for the nebular continuum level. Linear regressions to the respective data points are overlaid.}
\label{st_mode}
\end{figure*}

% The quest for unequivocal  
Figure \ref{st_mode} displays a comparison between the present-day stellar mass and mass-weighted stellar age determined by \starlight\ (blue points), \fado\ in its full consistency mode (orange points), and \fado$\rm _{ST}$ (purple points). This comparison clearly reveals an overestimation of the two physical properties when not considering the nebular continuum, which is more pronounced for the $\rm log_{10}$ $t_{\rm M}$ estimates by \starlight.
The right-hand side panels exhibit a positive correlation between the differences in estimated stellar mass and mass-weighted age when considering and not considering the nebular contribution and the logarithm of the flux resulting from the sum of the most prominent emission lines (namely, [OII] ($\lambda\lambda$ 3727, 3729), [OIII] ($\lambda\lambda$ 4363, 4959, 5007), H$\beta$, HeI ($\lambda\lambda$ 5876), [NII] ($\lambda\lambda$ 6548, 6584),  H$\alpha$, and [SII] ($\lambda\lambda$ 6717, 6731)), a direct proxy for the nebular continuum level. The over-plotted linear regressions clearly highlight the interdependence between the tested parameters, demonstrating that the higher the nebular contamination, the greater the stellar mass and mass-weighted age overestimation will be when using purely stellar codes (similar behaviour is also observed for luminosity-weighted age estimates). These findings are in overall agreement with the theoretical framework, as evidenced by Figs. 4 and 5 of \citet{Car19} where the authors perform a series of controlled tests on synthetic spectra with \fado\ and \starlight. Herein they demonstrate that, during evolutionary phases with a high sSFR, translating to an EW(H$\alpha$) $\ge$ 200\AA\ (such as the sampled EELGs which display an average EW(H$\alpha$) of $\sim$400\AA), \starlight\ tends to severely overestimate both the stellar mass and the mass-weighted stellar age estimates, whilst simultaneously strongly underestimating the light-weighted age and metallicity. \fado, on the other hand, is able to recover the mean stellar properties (i.e., total mass, mean age, and mean metallicity) with high accuracy ($\sim$0.2 dex). In addition, the theoretical work by \citet{Pap21} demonstrates that the overestimation of mass-weighted quantities observed in purely stellar codes such as \starlight\ and \steckmap\ \citep{Ocv11} relates to the inability of purely stellar models to fit the observed SED in the vicinity of the Balmer jump.
%where the authors demonstrate that \starlight, a purely stellar code, is unable to accurately estimate stellar mass and mass-weighted age in the presence of young stellar populations, which will significantly elevate the nebular continuum level.}

There is, however, a non-negligible disagreement between the difference in the mass-weighted stellar age as obtained by \fado$\rm _{ST}$\ and \fado, versus \starlight\ and \fado. Most likely, such reflects the significantly different optimisation algorithms employed by the two PSS codes (genetic differential evolution optimisation in the case of \fado\ and Markov chain Monte Carlo for \starlight), which translates into a complex behaviour enfolding several other factors, such as how the code engages the age--metallicity--extinction degeneracy (e.g., the PSS code might try to compensate for the nebular contribution not by selecting older SSPs but by increasing the stellar metallicity or intrinsic extinction). It is likewise impracticable to accurately predict in what manner a purely stellar code such as \starlight\ will fit the absorption stellar spectrum after being diluted by the nebular continuum. In any case, accurate evaluation of the role of the inclusion of the nebular contribution relative to the technical specifics of the fitting procedure (e.g., construction of the SSP library, the different mathematical recipes of the codes, in what manner the codes manage the age--metallicity--extinction degeneracy) requires an extensive series of tests on synthetic spectra, which on one hand is beyond the scope of this study and and on the other was previously assayed, as aforementioned.

%The observed discrepancy in the mass-weighted age estimates obtained by \fado$_{ST}$ and \starlight\ however must reflect a) the significantly different optimization algorithms employed by the two PSS codes (genetic differential evolution optimization in the case of \fado\ and Markov chain Monte Carlo for \starlight) and b) the well known age-metallicity-extinction.
%However, to unequivocally pinpoint a clear correlation between a possible age overestimation and nebular contamination through observational studies such as this one is not an easy task. 

% Through observational studies such as this one, it is likewise impracticable to accurately predict 
To note that there are additional problems which propagate in a non-linear manner, such as the excess at 1 Gyr which can be identified for both \fado\ modes, being more prominent for \fado\ in its full consistency mode. This effect, a known issue in spectral synthesis for which no solution currently exists \citep[see][and references therein]{Asari07} also prevents a direct and clear comparison. %In addition, the significantly different optimization algorithms employed by the two PSS codes (genetic differential evolution optimization in the case of \fado\ and Markov chain Monte Carlo for \starlight) must also play a role in the convergence for a different solution. Regardless, non-inclusion of a fundamental physical ingredient such as the nebular continuum which is irrefutably present in these spectra, can only result in inaccurate assessments.}

Furthermore, in view of the fact that both codes handle the emission lines in an entirely different fashion, an additional factor that could impact stellar-property determinations, and consequently affect the registered differences as obtained by the two spectral synthesis codes, is the treatment of the emission lines (i.e., in case of \starlight\ the emission lines must be masked, whereas \fado\ exploits some of them, in particular H$\alpha$ and H$\beta$ in addition to the Balmer and Paschen jumps, for the constraining of the nebular continuum contribution). However, both codes use the same set of emission lines, rendering it highly improbable that the origin of the discrepancies is the differences in the masking. Additionally, both codes auto-reject (faint) emission lines that might not be listed in the spectral mask. In the case where an emission line contains kinematically different components, or when an emission line comprises both a narrow and a broad (low-intensity) component, a single Gaussian fitting can be an issue, resulting in an inaccurate estimate of the stellar emission.
%a single Gaussian fitting can indeed be an issue when an emission line contains kinematically different  components, or when an emission line comprises both a narrow and a broad (low-intensity) component. 
In such circumstances, considering that the non-detection of the broad H$\alpha$ and H$\beta$ components would lead to a slight underestimation of the line flux, fitting a single Gaussian could underestimate the total flux, which would slightly influence the \fado\ fit, in its full self-consistency mode. However, visual inspection of our sample reveals a fraction of less than 10\% of galaxies with a possible broader, low-intensity component, and even then, a double-Gaussian fitting shows that the broad component contains less than 5\% of the total H$\alpha$ flux. Therefore, possible imperfections in the single-Gaussian fitting of lines cannot significantly account for the differences between \starlight, \fado\ and \fado$\rm _{ST}$. 

\begin{figure}[t]
\centering
\includegraphics[width=1\linewidth]{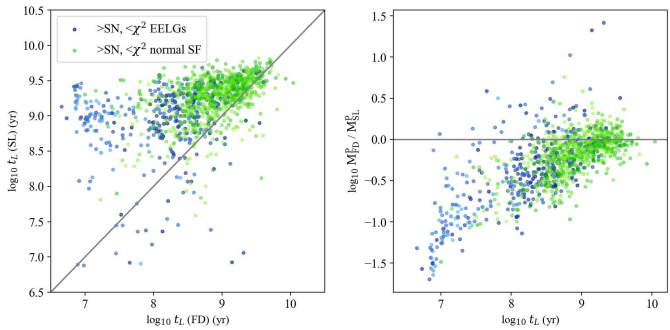}
\caption{Panel on the left-hand side displays the luminosity-weighted mean stellar age as obtained with \fado\ versus the same as obtained with \starlight. The right-hand side panel illustrates how the ratio between the obtained present-day stellar mass as obtained by \fado\ and \starlight\ correlates with the luminosity-weighted mean stellar age as obtained by \fado.}
\label{diffs}
\end{figure}

\begin{figure*}[b]
\centering
\includegraphics[width=1\linewidth]{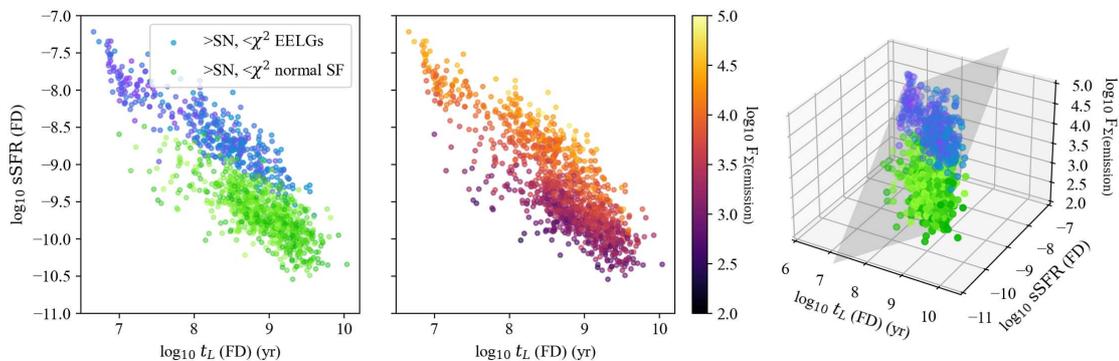}
\caption{Luminosity-weighted mean stellar age versus the logarithm of the sSFR for both galaxy samples (normal SF and EELGs), colour coded according to the different galaxy family (left panel) and according to the logarithm of sum of all the most prominent emission line fluxes (middle panel, units of 10$^{-16}$ erg/cm$^{2}$.s). The right panel illustrates the inter-connectivity of these three physical properties in a 3D parameter space, additionally displaying the resulting 3D surface that was fitted to the data.}
\label{age_sSFR_emiF}
\end{figure*}

Complementary, Fig. \ref{diffs} demonstrates that the lower the mean stellar age estimated by \fado, the higher the difference between both age and mass estimates, as obtained by both PSS codes. Not surprisingly, in light of the fact that younger (older) galaxies are expected to be less (more) massive, and as demonstrated by Figs. \ref{sSFR_FD_SL} and \ref{sSFR_FD_SL_normSF}, the mean stellar age derived by \fado\ is anti-correlated with the sSFR. Furthermore, the mean stellar age estimated by \fado\ is also related to the emission flux, that is, the sum of all the most prominent emission lines. Figure \ref{age_sSFR_emiF} documents these interrelationships: the left panel demonstrates how both galaxy families can be identified in the parameter space defined by the logarithm of the luminosity-weighted mean stellar age versus the logarithm of the sSFR, as given by \fado. Here two distinct linear relations are evident, displaced by a factor of $\sim$1 dex, however conserving approximately the same slope (-0.62 for the EELGs and -0.53 for the normal SF galaxies). The following panel evidences the underlying reason for the observed displacement. In effect, the sum of the flux of the most prominent emission lines is significantly contributing in shaping these relations. The inter-connectivity between the three galactic properties is depicted by the right-most panel, displaying a 3D surface modelling both data sets. The flat 3D surface is given by the equation:

%Categorically, the series of tests herein, in addition to the theoretical modelling by \citet{Car19} and \citet{Pap21} soundly point to the nebular continuum contribution as the main reason for the observed discrepancies between \starlight\ and \fado, in addition to the non-negligible effect of the individual mathematical recipes of the two PSS codes. Considering the aforementioned reasons and that it is not possible to attain a direct and clean comparison between the mean stellar ages provided by the two PSS codes or to ascertain how this is related to the nebular continuum level, we attempted to show this correlation indirectly.

% Categorically, the numerous trials performed herein evidences that} the sound determination of the stellar properties of SF galaxies requires a self-consistent approach such as the one taken by \fado, and that the analysis of the stellar content of SF galaxies with purely stellar codes should be avoided.

%\vspace{-0.8cm}
%Figure \ref{diffs} demonstrates that the lower the mean stellar age estimated by \fado, the higher the difference between both age and mass estimates, as obtained by both PSS codes. Not surprisingly, in light of the fact that younger (older) galaxies are expected to be less (more) massive, and as demonstrated by Figs. \ref{sSFR_FD_SL} and \ref{sSFR_FD_SL_normSF}, the mean stellar age derived by \fado\ is anti-correlated with the sSFR. Furthermore, the mean stellar age estimated by \fado\ is also related to the emission flux, that is, the sum of all the most prominent emission lines. Figure \ref{age_sSFR_emiF} documents these interrelationships: the left panel demonstrates how both galaxy families can be identified in the parameter space defined by the logarithm of the luminosity-weighted mean stellar age versus the logarithm of the sSFR, as given by \fado. Here two distinct linear relations are evident, displaced by a factor of $\sim$1 dex, however conserving approximately the same slope (-0.62 for the EELGs and -0.53 for the normal SF galaxies). The following panel evidences the underlying reason for the observed displacement: a third parameter, the sum of the flux of the most prominent emission lines, significantly contributes in shaping these relations. The inter-connectivity between the three galactic properties is depicted by the right-most panel, displaying a 3D surface modelling both data sets. The flat 3D surface is given by the equation:

\begin{equation}
 Z = 0.5656 \cdot X + 0.9846 \cdot Y + 7.8601
\end{equation}

where X, Y and Z correspond to the respective axis, representing the logarithm of the luminosity-weighted age estimated by \fado, the sSFR and the sum of the most prominent emission lines, respectively.

By employing this equation to predict the ages, given the values for the two additional quantities, and computing the differences between the predicted and observed ages, it is obtained a reasonable agreement, with a median of -0.02 and a RMS of 0.292 dex. The outcome from this simple exercise corroborates our finding of the critical role of the nebular emission in the sound determination of stellar properties derived from spectral synthesis, simultaneously acting as a prognostic instrument: on the basis of their sSFR and the level of their emission flux, the mean stellar age of SF galaxies can be approximately inferred. Additionally, by considering the relations displayed in Fig. \ref{diffs}, with these two physical properties it is possible to provide a rough estimate of the extent to which mass and age estimates from pure stellar codes such as \starlight\ should be adjusted to account for the nebular contribution.

Categorically, the series of tests herein, in addition to the theoretical modelling by \citet{Car19} and \citet{Pap21} soundly point to the nebular continuum contribution as the main reason for the observed discrepancies between \starlight\ and \fado, in addition to the non-negligible effect of the individual mathematical recipes of the two PSS codes.

\subsection{Comparison study between EELGs and normal SF galaxies}\label{res3}

%A detailed investigation of the stellar properties of EELGs is impracticable with standard PSS codes, which do not consider the substantial nebular contamination observed in these galaxies. Accordingly, in this work we employ instead an unconventional PSS code, \fado, which through a self-consistent approach allows to characterize the stellar content of galaxies. This procedure permitted to retrieve the stellar properties of a sample of $\sim$400 EELGs with available spectroscopic SDSS data, and to conduct a comparison study between $\sim$700 normal SF galaxies. 

The stellar properties of 414 EELGs and 697 normal SF galaxies were retrieved and compared. Generally we find that, for the same mass range, EELGs display, on average, systematically (but not substantially) lower mass- and luminosity-weighted ages. Unsurprisingly, however, the number of galaxies in each bin significantly varies depending on the galaxy family. 
%Stellar metallicities yield comparable values between both families.
Concerning the stellar surface density, in comparison to normal SF galaxies, EELGs are marginally less compact. These results are summarised in Table \ref{tab}.

\begin{table}[h]
\begin{tabular}{lllll}

 Log M$_{\star}$ (M$\odot$) & (6.5, 7.5{]} & (7.5, 8.5{]} & (8.5, 9.5{]} & (9.5, 10.5{]} \\
 \hline
nº of E                             & 0            & 36           & 322          & 56            \\
nº of nSF                           & 1            & 28           & 491          & 176           \\
log $\overline{t_{L}}$ E            & -            & 7.71         & 8.56         & 8.88          \\
log $\overline{t_{M}}$ E            & -            & 8.32         & 9.23         & 9.72          \\
log $\overline{t_{L}}$ nSF          & 7            & 8.31         & 9.05         & 9.28          \\
log $\overline{t_{M}}$ nSF          & 7            & 8.66         & 9.53         & 9.85          \\
log $\overline{\Sigma^{\star}}$ E   & -            & 7.11         & 7.76         & 8.23          \\
log $\overline{\Sigma^{\star}}$ nSF & 6.74         & 7.44         & 8.10         & 8.55          \\
%log $\overline{Z_{L}}$ E            & -            & 0.010        & 0.007        & 0.008         \\
%log $\overline{Z_{L}}$ nSF          & 0.007        & 0.008        & 0.007        & 0.007         \\
%log $\overline{Z_{M}}$ E            & -            & 0.009        & 0.007        & 0.006         \\
%log $\overline{Z_{M}}$ nSF          & 0.008     & 0.007        & 0.007        & 0.007                    
\end{tabular}
\caption{Table summarising the obtained mean values for both galaxy families (E for EELGs and nSF for normal SF galaxies) in four log of stellar mass bins, comprising information on, from top to bottom, number of galaxies, luminosity- and mass-weighted mean stellar age, and stellar surface density.}
\label{tab}
\end{table}

\begin{figure}[h]
\centering
\includegraphics[width=1\linewidth]{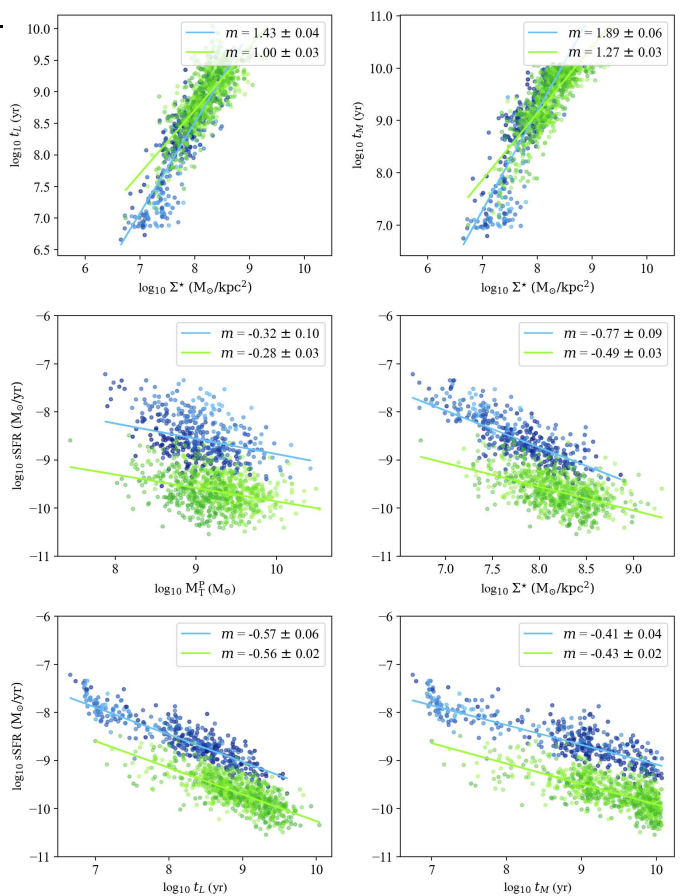}
\caption{Comparison of fundamental relations as obtained for EELGs (blue) and normal SF galaxies (green). The top row displays the correlations between mean stellar age (luminosity-weighted at the left-hand side and mass-weighted at the right-hand-side) and stellar surface density. The middle row contains, at the left-hand side, the relation between present day stellar mass and the logarithm of the sSFR and at the right-hand side, stellar surface density versus the logarithm of the sSFR. The left-hand side (right-hand side) of the bottom row displays the relations between the luminosity-weighted (mass-weighted) mean stellar age and the logarithm of the sSFR. The legends display the values of the slopes after performing linear fits to the respective data points and respective standard deviations}.
\label{compare_Age_SSD}
\end{figure}

The top row of Fig. \ref{compare_Age_SSD} evidences the contrast between EELGs and normal SF galaxies as obtained by processing the SDSS spectra with \fado, with respect to their compactness and mean stellar age (luminosity-weighted at the left-hand side and mass-weighted at the right-hand-side). Although there is a significant overlap, a linear fit to these relations reveals that EELGs display higher slopes. Nevertheless, such behaviour may not reflect a true disagreement between the nature of these two galaxy families, but instead the under-representativeness of normal SF galaxies with low mean ages and stellar surface densities. The left-hand side (right-hand side) of the middle row of the same figure reveals that, although there is a considerable displacement (i.e., for the same stellar mass or stellar surface density, EELGs yield sSFRs $\sim$1 dex higher), both galaxy families display similar slopes in the relation between stellar mass (stellar surface density) and the logarithm of sSFR. The same behaviour is visible in the lower panels, which display the relations between mean stellar age (luminosity- and mass-weighted) and the logarithm of the sSFR. 

In summary, comparison between EELGs and normal SF galaxies reveals that, despite the fact that EELGs consistently yield significant higher values of sSFR for the same stellar mass, no significant difference is seen in the mean ages of the galaxies of the two samples (the mean age of the EELGs is lower on average, but not substantially lower). This might occur due to the presence of a pronounced underlying old stellar component: despite the high values of sSFR, the contribution of the young stellar component is significantly lower than the same of the old. It might additionally reflect different SFHs: generally, normal SF galaxies may display more continuous SFHs, in contrast to EELGs which suddenly endure an extensive but brief star-formation episode. Detailed studies of high signal-to-noise-ratio integral field spectroscopy (IFS) data (which allows to spatially resolve the galaxy, thus revealing individual spectra of different morphological regions) with \fado\ might help to further disclose the nature of these galaxies and how they compare with normal SF galaxies.

\vspace{-0.5cm}
\section{Summary and Conclusions}\label{conc}

%A detailed investigation of the stellar properties of EELGs is impracticable with standard PSS codes, which do not consider the substantial nebular contamination observed in these galaxies. Accordingly, in this work we employ instead an unconventional PSS code, \fado, which through a self-consistent approach allows to characterize the stellar content of galaxies. This procedure permitted to retrieve the stellar properties of a sample of $\sim$400 EELGs with available spectroscopic SDSS data, and to conduct a comparison study between $\sim$700 normal SF galaxies. 

The nebular emission pertaining to young stellar populations that inhabit SF galaxies contaminates optical spectroscopic data, impeding the thorough characterisation of the stellar properties of the latter. The robust extraction of stellar properties in SF systems is only possible through an approach that self-consistently considers the nebular contribution, such as the one taken by \fado. 

In this work, we attempt to quantify differences in the stellar properties obtained using a purely stellar PSS code such as \starlight, versus \fado, which accounts for the nebular emission, for two set of galaxies: a sample of approximately 400 EELGs and a supplementary sample of approximately 700 normal SF galaxies. We confirm that familiar fundamental relations, such as stellar age versus stellar mass and stellar surface density, and sSFR versus stellar mass and stellar surface density and stellar age, can only be accurately inferred when accounting for the nebular contribution, yet again reflecting the importance of considering the nebular emission when characterising the stellar content of SF galaxies. This outcome is valid not only for EELGs but also for normal SF galaxies, demonstrating that even a mild nebular contribution is sufficient to substantially confuse purely stellar PSS codes, which systematically tend to retrieve higher stellar ages and masses. As anticipated, further exploration of the data has revealed that the two parameters that appear to regulate the observed differences between the stellar properties recovered through spectral synthesis are the sSFR and the emission flux, that is, the sum of the flux of the more prominent emission lines. 

Regarding the differences observed between EELGs and normal SF galaxies, the low-mass, low-stellar surface density locus of normal SF is underpopulated as compared to the EELG sample. For this region of the parameter space in particular, EELGs are significantly ($\sim$0.6 dex) younger than their normal SF counterparts. For higher mass bins, the stellar age differences are not as relevant as for the lower mass bins, despite the significantly higher sSFR. Such might be due to different SFHs: the recent star-forming activity in EELGs probably translates into a peaky SFH, while normal SF galaxies might endure SFHs of more continuous nature. High-quality IFS data, allowing to resolve stellar populations in time and space, might provide further insight in the near future.	

%Such can be ascertained 

%\nocite{*}

\begin{acknowledgements}
We thank the anonymous referee for valuable comments and suggestions.
I.B., J.V.M., J.I.P., C.K., E.P.M and A.A.P. acknowledge financial support from the State Agency for Research of the Spanish MCIU through the "Center of Excellence Severo Ochoa" award to the Instituto de Astrofísica de Andalucía (SEV-2017-0709). J.V.M., J.I.P., C.K., and E.P.M. acknowledge financial support from projects Estallidos6 AYA2016-79724-C4 (Spanish Ministerio de Economia y Competitividad), Estallidos7 PID2019-107408GB-C44 (Spanish Ministerio de Ciencia e Innovacion), and grant P18-FR-2664 (Junta de Andalucía). R.A. acknowledges support from ANID Fondecyt Regular 1202007.

%Additionally, we thank the EU for providing to Portugal a substantial fraction of the financial resources that allowed it to sustain a research infrastructure in astrophysics. Specifically, this work was carried out at an institute whose funding is provided to 85\% by the EU via the FCT (Funda\c{c}\~{a}o para a Ci\^encia e a Tecnologia) apparatus, through European and national funding via FEDER through COMPETE by the grants UID/FIS/04434/2013 \& POCI-01-0145-FEDER-007672 and PTDC/FIS-AST/3214/2012 \& FCOMP-01-0124-FEDER-029170. Additionally, this work was supported through FCT through the research grants [UID/FIS/04434/2019] UIDB/04434/2020 and UIDP/04434/2020. I.B. was supported by Instituto de Astrof\'isica e Ci\^encias do Espa\c{c}o through the research grant CIAAUP-30/2019-BID and by the FCT PhD::SPACE Doctoral Network (PD/00040/2012) through the fellowship PD/BD/52707/2014  funded by FCT (Portugal).
P.P. thanks Funda\c{c}\~{a}o para a Ci\^encia e a Tecnologia (FCT) for managing research funds graciously provided to Portugal by the EU and was supported through FCT grants UID/FIS/04434/2019, UIDB/04434/2020, UIDP/04434/2020 and the project "Identifying the Earliest Supermassive Black Holes with ALMA (IdEaS with ALMA)" (PTDC/FIS-AST/29245/2017).

This research has made use of the NASA/IPAC Extragalactic Database (NED) which is operated by the Jet Propulsion Laboratory, 
California Institute of Technology, under contract with the National Aeronautics and Space Administration.
\end{acknowledgements}

%\onecolumn
% References.tex
\bibliographystyle{apa}

%\clearpage

%\input{App.tex}

\end{document}